\documentclass[aps,pra,superscriptaddress,twocolumn,showpacs]{revtex4}%
\usepackage{amsfonts}
\usepackage{amsmath}
\usepackage{amssymb}
\usepackage{graphicx}

\begin{document}
\title{Hidden vortices in a Bose-Einstein condensate in a rotating double-well potential}
\author{Linghua Wen}
\affiliation{Institute of Physics, Chinese Academy of Sciences, Beijing 100190, China}
\affiliation{Department of Physics, Liaocheng University, Shandong 252059, China}
\author{Hongwei Xiong}
\affiliation{State Key Laboratory of Magnetic Resonance and Atomic and Molecular Physics,
Wuhan Institute of Physics and Mathematics, Chinese Academy of Sciences, Wuhan
430071, China}
\author{Biao Wu}
\affiliation{International Center for Quantum Materials, Peking University, Beijing 100871, China}
\affiliation{Institute of Physics, Chinese Academy of Sciences, Beijing 100190, China}
\date{\today}

\begin{abstract}
We study vortex formation in a Bose-Einstein condensate in a rotating
double-well potential. Besides the ordinary quantized vortices and elusive
ghost vortices, \textquotedblleft hidden\textquotedblright\ vortices are found
distributing along the central barrier. These hidden vortices are invisible
like ghost vortex but carry angular momentum. Moreover, their core size is not
given by the healing length, but is strongly influenced by the external
potential. We find that the Feynman's rule can be well satisfied only after
including the hidden vortices. There is no critical rotating frequency for the
formation of hidden vortex while there is one for the formation of ordinary
visible vortices. Hidden vortices can be revealed in the free expansion of the
Bose-Einstein condensates. In addition, the hidden vortices in a Bose-Einstein
condensate can appear in other external potentials, such as a rotating
anisotropic toroidal trap.

\end{abstract}

\maketitle

\section{Introduction}

Double-well (DW) potential is an important model potential for its simplicity
and yet richness in physics. The properties of Bose-Einstein condensates
(BECs) in a DW potential have been studied extensively since the realization
of BECs \cite{Smerzi}. With the advances of technology, DW potentials for
ultracold atoms can now be realized in experiments with great controllability
and precision \cite{Albiez,Hofferberth1,Esteve}. One particular interesting
development is the possibility to rotate a DW potential via radiofrequency
dressing \cite{Hofferberth2}. This opens the door to study the behavior of
vortices of degenerate quantum gas in a rotating DW potential. As topological
defect, quantized vortices have contributed greatly to reveal the phase
coherence, superfluidity and nonlinear phenomena for degenerate quantum gases,
and have been the subject of extensive experimental and theoretical studies
\cite{Lin,Spielman,Madison,Abo-Shaeer,Haljan,Hodby,Zwierlein,Weiler,Fetter,Cooper,Dalfovo,Stringari,Penckwitt,Baym,Aftalion1,Saito,Ostrovskaya}%
.

The DW potential also offers a unique testing ground for the well-known
Feynman's rule of vortex \cite{Feynman}. The Feynman's rule is very powerful
relation that links the total number of vortices with rotation angular
frequency. Feynman argued that for a rotating superfluid with angular
frequency $\Omega$, the superfluid should be regarded as a classical fluid
when it reaches the steady state. This leads to an important mathematical
relation that the total number of vortices $N_{v}$ in an area $A$ is linearly
proportional to $\Omega$, $2\pi\hbar N_{v}/m=2\Omega A$. Alternatively, the
Feynman's rule can expressed as $l_{z}/\hbar=N_{v}/2$ with $l_{z}$ being the
mean angular momentum per atom at equilibrium
\cite{Fetter,Cooper,Tsubota,Feynman}. This rule was originally for a uniform
superfluid helium, and has been intensively studied both theoretically
\cite{Tsubota,Feder} and experimentally \cite{Abo-Shaeer,Haljan} for a BEC
trapped in single harmonic potential. It is interesting to know how the rule
fares in a more complicated geometric confinement. The DW potential provides a
clear opportunity to answer this question.

In this paper we have conducted a comprehensive two-dimensional (2D) numerical
study of vortex formation in a BEC in a rotating DW potential. We find
surprisingly that the \textit{in situ} density distribution seems to violate
significantly the Feynman's rule in that the total number of vortices with
visible core is significantly smaller than $2l_{z}/\hbar$. With the belief
that the Feynman's rule should hold in some form, we carefully analyze the
results and find these \textquotedblleft missing" vortices are distributed
along the central barrier of the DW potential. Unlike the usual vortex, these
vortices have no visible cores but have phase singularities, and their core
size is not given by the healing length but is strongly influenced by the
external potential. For this reason, they may be called \textquotedblleft
hidden vortices\textquotedblright. With the inclusion of the hidden vortices,
one recovers the Feynman's rule.

These hidden vortices remind us of ghost vortices found numerically in Ref.
\cite{Tsubota}. The ghost vortices always lie at the outskirts of the
condensate where the particle density $\left\vert \psi\right\vert ^{2}$ is
very small. As a result, similar to hidden vortex, they show up in numerical
results as phase singularities but have no visible cores. However, there are
key differences: the ghost vortex carries no angular momentum while hidden
vortex does; the core size of ghost vortex is determined by healing length
like the usual visible vortex while that of hidden vortex by the shape of
external potential. In addition, we find that with the increasing of rotation
frequency $\Omega$ the hidden vortices appear first in the DW system, followed
by ghost vortices and usual visible vortices. Furthermore, the angular
momentum can be put gradually into the BEC via the generation of the hidden
vortices while the emergence of a visible vortex is still accompanied by a
jump in the system angular momentum. Although the hidden vortices are
invisible in the \textit{in situ} density distribution, after free expansion
of the BEC, they can appear in the density distribution because of their
stable topological structure.

It is well known that there exists a special type of vortex called Josephson
vortex (or fluxon) in a long superconducting Josephson junction \cite{Barone}
or between two weakly-coupled BECs \cite{Kaurov,Brand}. These Josephson
vortices can be regarded as hidden vortices. However, hidden vortex is a more
general notion than the Josephson vortex (fluxon) as hidden vortex can exist
in non-DW potential (or non-Josephson structure). We have used rotating
anisotropic toroidal trap to illustrate this point.

This paper is organized as follows. In Sec. II, we present a phenomenological
model to describe the dynamics of a BEC confined in a rotating DW potential in
the presence of dissipation. Hidden vortices are found in the rotating DW BEC,
where the Feynman's rule is satisfied only after including these hidden
vortices. A simple and feasible scheme is proposed to observe the hidden
vortices. In Sec. III, we study the vortex formation process and the critical
rotating frequency in the rotating DW BEC. In Sec. IV, we discuss the hidden
vortices in a BEC confined in a rotating anisotropic toroidal trap. Sec. V
provides a summary and discussion.

\section{Hidden vortices in a rotating DW BEC and Feynman's rule}

We consider the situation where the condensate is tightly confined in the
axial direction ($\omega_{x},\omega_{y}\ll\omega_{z}$) so that the system is
effectively two dimensional. The DW potential is described by
\begin{equation}
V_{dw}(x,y)=\frac{x^{2}+\lambda^{2}y^{2}}{4}+V_{0}e^{-x^{2}/2\sigma^{2}%
}\,,\label{double well}%
\end{equation}
where $V_{0}$ and $\sigma$ are the height and width of the potential barrier,
respectively, and $\lambda=\omega_{y}/\omega_{x}$ denotes the anisotropy
parameter of the harmonic trap. In the presence of dissipation, the order
parameter in the frame rotating with the angular velocity $\Omega$ around the
$z$ axis obeys the time-dependent Gross-Pitaevskii (GP) equation
\begin{equation}
(i-\gamma)\frac{\partial\psi}{\partial t}=\left[  -(\nabla_{x}^{2}+\nabla
_{y}^{2})+V_{dw}+c\left\vert \psi\right\vert ^{2}-\Omega L_{z}\right]
\psi.\label{GPE}%
\end{equation}
Here $L_{z}=i(y\partial_{x}-x\partial_{y})$ is the $z$ component of the
angular-momentum operator, $\gamma$ characterizes the degree of dissipation,
and $c$ is the 2D interatomic interaction strength. In this work, length,
time, energy, angular momentum, and rotation angular frequency are in units of
$d_{0}=\sqrt{\hbar/2m\omega_{x}}$, $1/\omega_{x}$, $\hbar\omega_{x}$, $\hbar$,
and $\omega_{x}$, respectively. The phenomenological dissipation model
(\ref{GPE}) is a variation of that in Ref. \cite{Tsubota}. For the case of a
BEC in a rotating harmonic trap, our computation results agree well with the
experimental observations in Ref. \cite{Madison} and the simulation results in
Ref. \cite{Tsubota}.

In our calculations, we first obtain the initial ground-state order parameter
in the DW potential by imaginary time propagation method \cite{Wu,Liu,Wen} for
$\Omega=0$. The vortex formation process is then studied by solving
numerically Eq. (\ref{GPE}) with different $\Omega$. Here we consider the BECs
of $^{87}$Rb atoms with repulsive interaction. The system parameters are
chosen to be $\omega_{x}=\omega_{y}=2\pi\times40$ \textrm{Hz}, $\omega
_{z}=2\pi\times800$ \textrm{Hz}, $V_{0}=40$, $\sigma=0.5$, $c=600$. In Eq.
(\ref{GPE}), the variation of $\gamma$ only influences the relaxation time
scale but does not change the dynamics of vortex formation and the ultimate
steady vortex structure. In our computation, we choose $\gamma=0.03$, which
corresponds to a temperature of about $0.1T_{c}$ \cite{Choi}.

Fig. 1(a) shows the steady density distribution $\left\vert \psi\right\vert
^{2}$ at $t=250$ for the DW potential rotating with $\Omega=0.9$. From this
\textit{in situ} density distribution, we see a pair of ordinary vortex
lattices with triangular structure as expected from the rotating DW
configuration. However, by simply counting, we find something surprising. In
Fig. 1(a), the total number of vortices is $N_{v}=18$. Numerical results show
that the mean angular momentum per atom $l_{z}=\int\int\psi^{\ast}L_{z}\psi
dxdy/\int\int\left\vert \psi\right\vert ^{2}dxdy$ is about $l_{z}\approx16\gg
N_{v}/2$. It seems that the Feynman's rule
\cite{Fetter,Cooper,Tsubota,Feynman} is no longer satisfied. For other
rotation frequencies, this seemingly significant violation of Feynman's rule
is also found. With the belief that the Feynman's rule should always hold,
this violation indicates that some angular momentum is missing and is not
manifested in the form of ordinary vortices.

%%%%%%%%%%%%%%%%%%%%%%%%%%%%%%%%%%%%
\begin{figure}[ptb]
\centerline{\includegraphics*[width=8.2cm]{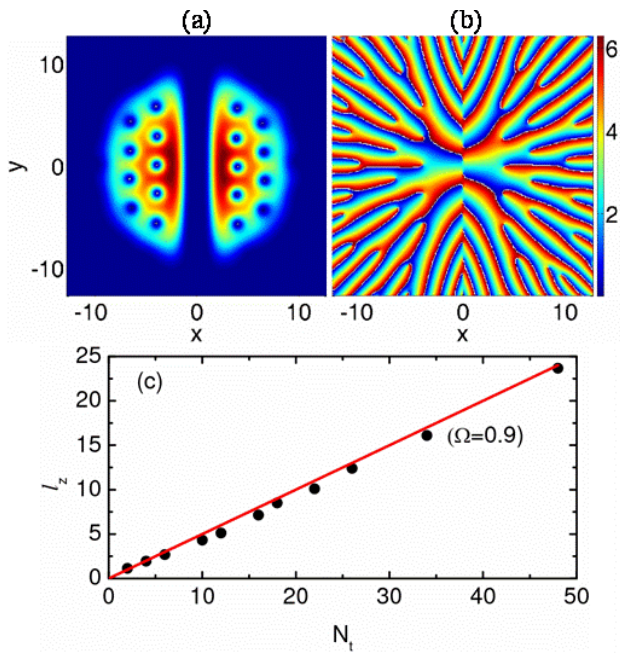}}\caption{(color
online) (a) Density distribution $\left\vert \psi\right\vert ^{2}$ and (b)
phase distribution of $\psi$ at $t=250$ after rotating the system with
$\Omega=0.9$. (c) $l_{z}$ versus $N_{t}/2$ for different $\Omega$, where the
line is $l_{z}=N_{t}/2=$ $\left(  N_{v}+N_{h}\right)  /2$ and the solid
circles denote the numerical results. The value of the phase varies
continuously from $0$ to $2\pi$. The darker color area indicates the lower
density or phase. The small derivation between the line and the circles in (c)
may be due to the inhomogeneous condensate density \cite{Tsubota}.}%
\label{Fig1}%
\end{figure}
%%%%%%%%%%%%%%%%%%%%%%%%%%%%%%%%%%%%

%\begin{figure}[htbp]
%\includegraphics[width=6.8cm]{Fig1.eps}
%\caption{(color online) (a) Density distribution $\left\vert \protect\psi %
%\right\vert ^{2}$ and (b) phase distribution of $\protect\psi $ at
%$t=250$
%after rotating the system with $\Omega =0.9$. (c) $l_{z}$ versus $N_{t}/2=$ $%
%\left( N_{v}+N_{h}\right) /2$ for different $\Omega $, where the line is $%
%l_{z}=N_{t}/2$ and the solid circles denote the numerical results.
%The value of the phase varies continuously from $0$ to $2\protect\pi
%$. The darker color area indicates the lower density or phase. The
%small derivation between the line and the circles in (c) may be due
%to the inhomogeneous condensate density \protect\cite{Tsubota}.}
%\end{figure}

To find the missing angular momentum, we look into the phase distribution of
$\psi(x,y,t=250)$, which is plotted in Fig. 1(b). We find that besides the
phase singularities corresponding to the above mentioned vortices, there are
other phase defects, distributed along the central barrier and the outskirts
of the cloud. The initial reaction is that these phase defects, which are
invisible in the \textit{in situ }density distribution, are ghost vortices
discussed in Ref. \cite{Tsubota}. Since ghost vortices are known not carrying
angular momentum, it seems that these invisible phase defects can not account
for the missing angular momentum. However, a more careful examination shows otherwise.

The phase singularities along the central barrier are not ghost vortices and
they carry angular momentum. To see this, we assume $N_{h}$ is the total
number of phase singularities along the central barrier, and $N_{t}%
=(N_{v}+N_{h})$ is the sum of $N_{v}$ and $N_{h}$. From Fig. 1(b), we have
$N_{h}=16$. If we include them, the Feynman's rule $l_{z}\simeq N_{t}%
/2=(N_{v}+N_{h})/2$ is well satisfied, indicating that these phase
singularities do carry angular momentum like the usual visible vortices. We
have checked other rotating frequencies and that the Feynman's rule can always
be well satisfied by including the phase defects along the central barrier. In
Fig. 1(c), we have plotted the dependence of $l_{z}$ on $N_{t}/2$ for
different rotation angular frequencies. The solid line is the Feynman's rule
$l_{z}=N_{t}/2$ while the solid circles are the numerical results of $l_{z}$
at equilibrium. The excellent agreement between them strongly support that
phase singularities along the central barrier carry angular momentum, and thus
are not ghost vortices. We call these topological defects along the central
barrier \textquotedblleft hidden vortices\textquotedblright\ for their
difference from the ordinary vortex and the ghost vortex.

Two factors are involved in why a hidden vortex carries angular momentum while
a ghost vortex does not: location and core size. To see this, we consider a
phase defect with a singly quantized circulation in the condensate. The
angular momentum carried by this phase defect varies with the location as
$l_{z}\sim(1-r^{2}/R^{2})$ \cite{Fetter}, where $R$ is the size of the
condensate and $r$ is the distance from the center. Since a ghost vortex
always lies on the outskirts of the condensate, meaning $r\sim R $, its
contribution to angular momentum is negligible. For a hidden vortex, which
locates near the center of the condensate, we have $r<R$. Therefore, its
contribution to the angular momentum is significant and needs to be counted
for the Feynman's rule. Furthermore, the core size of a non-hidden vortex is
about the healing length $\xi=\sqrt{1/nc}$ ($n$ is the local density of the
condensate without vortices). Since $n$ is very small for a ghost vortex, the
core size of a ghost vortex approaches an infinite value. As a result, ghost
vortices neither contribute to the angular momentum nor the energy of the
system. For a hidden vortex, $n $ is also very small at its location. However,
its core size is determined by the barrier width, not the healing length.
Therefore, hidden vortex can contribute to the angular momentum. With the
local density approximation, our numerical calculations do show that the
hidden vortices carry significant angular momentum while the angular momentum
due to a ghost vortex can be negligible.

%\section{A feasible scheme to observe hidden vortices}

Even though the hidden vortices are invisible in the \textit{in situ} density
distribution as shown in Fig. 1(a), we find numerically that they show up in
the cloud after free expansion (see the following discussion). This makes it
possible to observe and test the existence of these hidden vortices experimentally.

%Note that, in contrast, there is no feasible way at
%present to observe a ghost vortex.

%To simulate the free expansion of the BEC after switching off the DW
%potential at the end of rotation, we use the laboratory frame and solve the
%dynamic equation%
%\begin{equation}
%i\frac{\partial}{\partial\tau}\varphi(x,y,\tau)=\left[ -(\nabla_{x}^{2}+%
%\nabla_{y}^{2})+c\left\vert \varphi\right\vert ^{2}\right] \varphi(x,y,\tau),
%\label{free expansion}
%\end{equation}
%where the initial state $\varphi(x,y,\tau=0)=$ $\psi(x,y,t)$ is the solution
%of Eq. (\ref{GPE}) at the end of the rotation. In Eq. (\ref{free expansion}%
%), we denote the expansion time with symbol $\tau$ in order to discriminate
%it from $t$. Since this expansion time scale is short, the dissipation can
%be safely omitted in Eq. (\ref{free expansion}). In fact, our simulation
%shows that the inclusion of damping during the expansion only influences the
%lifetime of the vortices.

We use the state shown in Figs. 1(a)-1(b) as an example. After a short
expansion time, the state begins to look very differently. In Fig. 2, the
density distribution and phase distribution at the expansion time $\tau=4$ is
plotted. We see clearly in Figs. 2(a)-2(b) that, besides eighteen vortices
already shown in the \textit{in situ} density distribution [see Fig. 1(a)], a
series of new ordinary vortices appear along the symmetric axis of two
condensates. These new visible vortices originate physically from the hidden
vortices. It is not difficult to understand the revelation of the hidden
vortices during the free expansion. The core of a vortex (hidden, ghost, or
visible) is also a velocity singularity, where the velocity approaches
infinity. Because the kinetic energy should be finite, during the free
expansion where the angular momentum is conserved, no atoms will be allowed
into the core area. As a result, the core is stable and will not be destroyed.
At the same time, as the two BECs begin to overlap, atoms begin to move into
the central region and fill the space between hidden vortex cores, rendering
them visible. No all hidden vortices can be revealed during free expansion. As
shown in Fig. 2(c), there still exist several remnant hidden vortices along
the boundary between the two flows [see Fig. 2(c)], which is due to the strong
repulsion and pushing of the newly formed visible vortices during the
expansion. Note that in our simulation, we did not find the generation of the
vortex-antivortex\ pairs predicted in the numerical study of the interference
of sliced BECs without rotation \cite{Carretero}.

Because the ghost vortices lie on the outskirts, the locations of their phase
defects move outward during the free expansion. Consequently, the density in
the regime of the ghost vortices is always negligible. This is the reason that
the ghost vortices will not become visible vortices during the expansion.

%%%%%%%%%%%%%%%%%%%%%%%%%%%%%%%%%%%%
\begin{figure}[ptb]
\centerline{\includegraphics*[width=8.3cm]{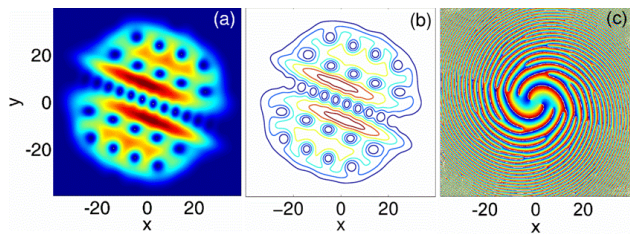}}\caption{(color
online) (a) Density distribution, (b) density contour, and (c) phase
distribution after the cloud free expands for $\tau=4$. Before the free
expansion, the system has rotated with $\Omega=0.9$ for $t=250$. The darker
color area or contour indicate the lower density or phase. }%
\label{Fig2}%
\end{figure}
%%%%%%%%%%%%%%%%%%%%%%%%%%%%%%%%%%%%

%\begin{figure}[tbp]
%\includegraphics[width=8.3cm]{Fig2.eps}
%\caption{(color online) (a) Density distribution, (b) density
%contour, and (c) phase distribution after the cloud free expands for
%$\protect\tau =4$.
%Before the free expansion, the system has rotated with $\Omega =0.9$ for $%
%t=250$. The darker color area or contour indicate the lower density
%or phase. }
%\end{figure}

\section{Vortex formation process and critical rotating frequency for a
rotating DW BEC}

The vortex formation process with the rotating DW potential is drastically
different from the one with a single harmonic potential. There is a critical
angular frequency for a rotating single-well potential to create vortices:
when the angular frequency $\Omega$ is below the critical angular frequency,
only ghost vortices are formed at the outskirts of the BEC cloud and the cloud
does not carry any significant angular momentum; when $\Omega$ is larger than
the critical frequency, visible vortices begin to appear along with a jump in
the angular momentum \cite{Tsubota}. For the rotating DW system, the vortex
formation starts with a pair of hidden vortices. As seen in Figs. 3(a)-3(e),
the hidden vortex pair begin their formation at the ends of the potential
barrier, then move towards the center. This is followed by a sequence of other
hidden vortex pairs. Ghost vortices begin to appear only after several pairs
of hidden vortices are already formed. Eventually at the critical rotating
frequency $\Omega_{c}=0.59$, a pair of ordinary visible vortices are formed
[see Fig. 3(f)], along with a jump in the system angular momentum [see Fig. 3(g)].

The dependence of the angular momentum per atom $l_{z}$ on the rotating
frequency $\Omega$ is shown in Fig. 3(g). It is clear from the figure that the
angular momentum $l_{z}$ increases gradually and continuously with $\Omega$
until a jump occurs at $\Omega_{c}=0.59$.\ Along with Figs. 3(a)-3(e), this
means that the hidden vortices can gradually increase the system angular
momentum as they move towards the center. The ghost vortex has no capacity to
carry angular momentum. As demonstrated clearly in Figs. 3(d)-3(e), even as a
pair of ghost vortices move towards the center and eventually become a pair of
ordinary visible vortices, the change in the angular momentum is very sudden
as witnessed by the jump in Fig. 3(g).

%%%%%%%%%%%%%%%%%%%%%%%%%%%%%%%%%%%%
\begin{figure}[ptb]
\centerline{\includegraphics*[width=8.2cm]{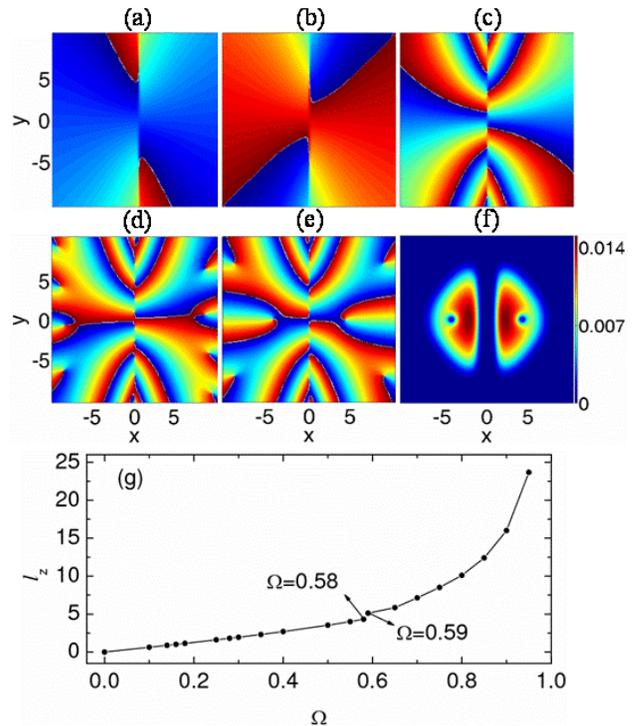}}\caption{(color
online) Phase distributions of $\psi$ at $t=250$ for different rotating
frequencies: (a) $\Omega=0.1$; (b) $\Omega=0.12$; (c) $\Omega=0.5$; (d)
$\Omega=0.58$; (e) $\Omega=0.59$, respectively. (f) The density distribution
$\left\vert \psi\right\vert ^{2}$ at $t=250 $ for $\Omega=0.59$. (g) Angular
momentum per atom $l_{z}$ versus $\Omega$. The darker color area indicates the
lower phase or density. }%
\label{Fig3}%
\end{figure}
%%%%%%%%%%%%%%%%%%%%%%%%%%%%%%%%%%%%

%\begin{figure}[tbp]
%\includegraphics[width=8.2cm]{Fig3.eps}
%\caption{(color online) Phase distributions of $\protect\psi $ at
%$t=250$ for different rotating frequencies: (a) $\Omega =0.1$; (b)
%$\Omega =0.12$; (c) $\Omega =0.5$; (d) $\Omega =0.58$; (e) $\Omega
%=0.59$, respectively. (f)
%The density distribution $\left\vert \protect\psi \right\vert ^{2}$ at $%
%t=250 $ for $\Omega =0.59$. (g) Angular momentum per atom $l_{z}$ versus $%
%\Omega $. The darker color area indicates the lower phase or
%density.}
%\end{figure}

In virtue of data fit, another interesting feature in Fig. 3(g) is that, for
$\Omega<\Omega_{c}$, $l_{z}$ increases linearly; in contrast, for $\Omega
\geq\Omega_{c}$, $l_{z}$ grows exponentially. These two different regimes
marked by well linear growth and perfectly exponential growth are likely
associated with the fact that the hidden vortices only form along the central
barrier but the ordinary visible vortices can emerge in the whole regions of
the cloud.

\section{Hidden vortices in a BEC confined in a rotating toroidal trap}

Hidden vortex can exist in non-DW potentials. To illustrate this point, we
consider a BEC confined in a rotating toroidal trap. The toroidal trap is
given by
\begin{equation}
V_{tt}(x,y)=\frac{x^{2}+y^{2}}{4}+V_{0}e^{-(\alpha x^{2}+y^{2}/\alpha
)/2\sigma^{2}}\,,\label{toroidal trap}%
\end{equation}
where $\alpha$ characterizes the anisotropy of the 2D central barrier in the
toroidal trap. $\alpha=1$ corresponds to a circular toroidal trap, which has
recently been studied by Aftalion \textit{et al.}\cite{Aftalion2}. We focus on
the anisotropic (deformed) toroidal trap ($\alpha\neq1$), where the lack of
the rotation symmetry excludes the formation possibility of multi-quantized
vortex (giant vortex) in the center of the trap for sufficiently large
rotation frequency $\Omega$ and sufficiently narrow barrier. The numerical
procedure for this toroidal trap is identical to the DW potential.
%In the rotating frame, the GP equation reads%
%\begin{equation}
%(i-\gamma )\frac{\partial \psi }{\partial t}=\left[ -(\nabla _{x}^{2}+\nabla
%_{y}^{2})+V_{tt}+c\left\vert \psi \right\vert ^{2}-\Omega L_{z}\right] \psi .
%\label{dynamic equation in toroidal trap}
%\end{equation}
In Fig. 4 we display the steady density distributions $\left\vert
\psi\right\vert ^{2}$ (left) and the corresponding phase distributions of
$\psi$ (right) at $t=250$ for the anisotropic toroidal trap rotating with
$\Omega=0.6$ (top) and $\Omega=0.9$ (bottom), respectively. The parameters are
$V_{0}=40$, $\alpha=0.8$, $\sigma=1$, $\gamma=0.03$, and $c=600$.

At $\Omega=0.6$, two visible vortices appear in the \textit{in situ} density
distribution as shown in Fig. 4(a). Moreover, there is an ellipsoid density
hole in the trap center which looks like a giant vortex. In the phase
distribution displayed in Fig. 4(b), we see that besides two phase
singularities corresponding to the two visible vortices, there are other phase
defects, which are distributed along the long axis of the central barrier and
at the outskirts of the cloud. The four singly quantized phase defects along
the long axis of the central barrier show that the ellipsoid density hole is
not a giant vortex. Our numerical simulation further indicates that the four
single-quantized phase defects carry angular momentum and satisfy the
Feynman's rule together with the two visible vortices. Therefore, they are
four singly quantized hidden vortices. On the other hand, the phase defects at
the outskirts of the cloud are ghost vortices because they contribute no
angular momentum to the system.

With the increase of rotation frequency, more vortices nucleate and a
triangular vortex lattice forms eventually [see Fig. 4(c)]. At the same time,
as shown in Fig. 4(d), more hidden vortices also show up in the central
barrier region, e.g., there are six hidden vortices for $\Omega=0.9$. However,
these hidden vortices do not form triangular lattice, and they are still
distributed along the long axis of the ellipsoid central barrier. This shows
from another perspective that the hidden vortex is different from the usual
visible vortex. The local enlargements of Figs. 4(b) and 4(d) are given in
Fig. 5, where the circles denote the positions of hidden vortices.

%%%%%%%%%%%%%%%%%%%%%%%%%%%%%%%%%%%%
\begin{figure}[ptb]
\centerline{\includegraphics*[width=7.4cm]{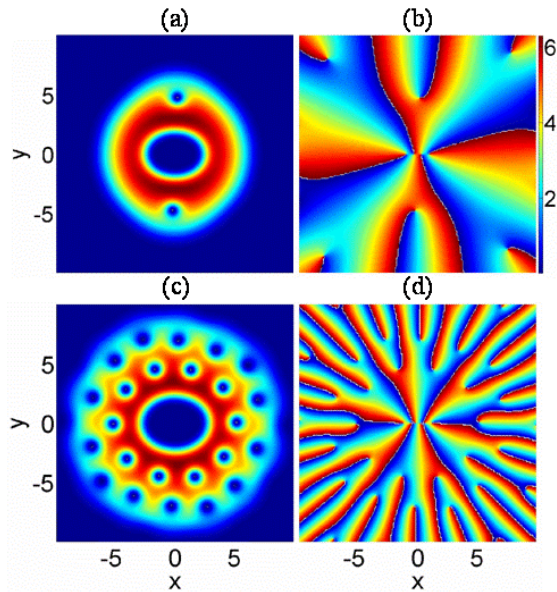}}\caption{(color
online) Density distributions $\left\vert \psi\right\vert ^{2}$ (left) and
phase distributions of $\psi$ (right) at $t=250$ for the toroidal trap
rotating with $\Omega=0.6$ (top) and $\Omega=0.9$ (bottom), respectively. The
corresponding parameters are $V_{0}=40$, $\alpha=0.8$, $\sigma=1$,
$\gamma=0.03$, and $c=600$. The value of the phase varies continuously from
$0$ to $2\pi$. The darker color area indicates the lower density or phase. }%
\label{Fig4}%
\end{figure}
%%%%%%%%%%%%%%%%%%%%%%%%%%%%%%%%%%%%

%%%%%%%%%%%%%%%%%%%%%%%%%%%%%%%%%%%%
\begin{figure}[ptb]
\centerline{\includegraphics*[width=7.6cm]{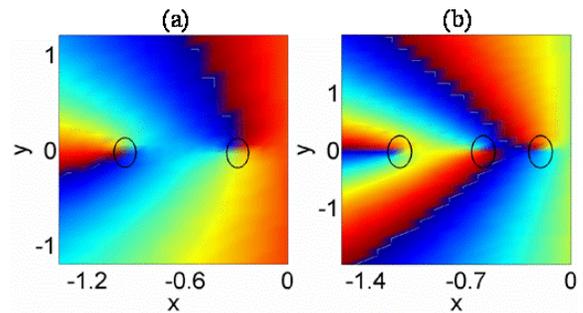}}\caption{(color
online) (a) Local enlargement of Fig. 4(b) and (b) that of Fig. 4(d), where
the circles mark the positions of hidden vortices.}%
\label{Fig5}%
\end{figure}
%%%%%%%%%%%%%%%%%%%%%%%%%%%%%%%%%%%%

\section{Discussion and Summary}

In summary, we have investigated numerically the formation of vortices in a
rotating double-well potential. We found that, other than usual visible vortex
and ghost vortex, there exists another type of vortex, which we call hidden
vortex. Unlike the usual visible vortex, these hidden vortices are invisible
in the \textit{in situ} density distribution. They differ also from the ghost
vortex by being able to carry angular momentum. In addition, the core size of
hidden vortex is not given by the healing length, and is strongly influenced
by the shape of external potential. Only after including the hidden vortices
can the Feynman's rule be satisfied.

Hidden vortex has appeared in literature in other names. For example, the
magnetic fluxons in a superconducting long Josephson junction in a parallel
magnetic field \cite{Barone}, Josephson vortices between two long parallel
coupled atomic BECs \cite{Kaurov}, and rotational fluxons of BECs in rotating
coplanar double-ring traps \cite{Brand}. The giant vortices (or, sometimes,
called \textquotedblleft phantom vortices\textquotedblright) in a cylindrical
hard-walled bucket or a quadratic plus quartic trap \cite{Fischer} or a
circular toroidal trap \cite{Aftalion2} can also be regarded as a form of
hidden vortex. However, as we have illustrated in the last section with an
anisotropic toroidal trap, hidden vortex can occur in many settings other than
previously mentioned structure or potentials. Therefore, hidden vortex is a
more general notion that encompass all the essential features of Josephson
vortex and giant vortex. At the same time, these names, such as Josephson
vortex and giant vortex, each coined for a special potential, show that it is
necessary to distinguish hidden vortex from the usual visible vortex and the
ghost vortex.

\begin{acknowledgments}
We thank B. L. Lv, K. J. Jiang, Y. P. Zhang and L. Mao for helpful
discussions. This work was supported by the NSFC (10825417, 10875165,
10847143), the NSF of Shandong Province (Q2007A01), and Ph.D. Foundation of
Liaocheng University.
\end{acknowledgments}

\end{document}